\documentclass[12pt]{article}
\title{
 The  static $Q\bar Q$ interaction  at small distances
 and OPE violating terms.}
 \author{Yu.A.Simonov\\Institute of Theoretical and Experimental
Physics}
 \date{}
\newcommand{\be}{\begin{equation}}
\newcommand{\ee}{\end{equation}}
\def\la{\mathrel{\mathpalette\fun <}}

\def\fun#1#2{\lower3.6pt\vbox{\baselineskip0pt\lineskip.9pt
\ialign{$\mathsurround=0pt#1\hfil
##\hfil$\crcr#2\crcr\sim\crcr}}}

\begin{document}
\maketitle

\begin{abstract}

Nonperturbative contribution to the one-gluon exchange produces a
universal linear term in the static potential at small distances
$\Delta V=\frac{6N_c \alpha_s \sigma r}{2\pi}$. Its role in the
resolution of long--standing discrepancies in the  fine splitting of
heavy quarkonia and improved agreement with lattice data for static
potentials is discussed, as well as implications for OPE violating
terms in other processes.
\end{abstract}

 1. Possible nonperturbative contributions from small distances have
 drawn a lot of attention recently [1,2]. In terms of interquark
 potential
  the appearance of linear terms in the static potential $V(r) =$
 const $r$, where $r$ is  the distance between charges, implies
 violation of OPE, since const $\sim $ (mass)$^2$ and this dimension
 is not available in terms of field operators.  There is however some
 analytic [1,2] and numerical evidence [3] for the possible existence
 of such terms $O(m^2/Q^2)$ in asymptotic expansion at large $Q$.

On a more phenomenological side the presence of linear term at small
distances, $r<T_g$, where $T_g$ is the gluonic correlation length
 [4,5], is required by at least two sets of data.

First, the detailed lattice data [6] do not  support
much weaker  quadratic behaviour of  $V(r)\sim$ const $r^2$,
following from OPE and field correlator method [4,5], and instead
prefer the same  linear form $V(r) =\sigma r$ at all distances (in
addition  to perturbative $-\frac{ C_2\alpha_s}{r}$ term). Second,
the small--distance linear term is necessary for  the description of
the fine  splitting in heavy quarkonia, since the spin--orbit Thomas
term $V_t=-\frac{1}{2m^2r}\frac{dV}{dr}$ is sensitive to the small
$r$ region and additional linear contribution at $r<T_g$ is needed to
fit the experimental splitting [7]. Moreover lattice calculations
[8] display the  $1/r$ behaviour of $V_t$
 in all measured region up
to $r=0.1 fm$.

 Of crucial importance is the sign of the $O(m^2/Q^2)$ term, since
 the usual screening correction (real $m$) leads to negative sign of
 linear potential, and one needs small--distance nonperturbative (NP)
 dynamics, which produces negative (tachyonic) sign of $m^2$ [1,2].
 Phenomenological implications of such contributions have been
 thouroughly studied in[1]. In what follows we show that interaction
 of gluon spin with NP background indeed yields tachyonic gluon mass
 at small distances.

2. In this letter we report the first application of the
systematic background perturbation theory [9,10] to the problem in
question.  One starts with the decomposition of the full gluon vector
potential $A_\mu$ into nonperturbative  (NP) background $B_\mu$ and
perturbative field $a_\mu$,
\be A_\mu=B_\mu+a_\mu, \ee
 and make use of  the
'tHooft identity for the  partition function
\be
Z=\int DA_\mu
e^{-S(A)}= \frac{1}{N} \int DB_\mu\eta(B) \int Da_\mu e^{-S(B+a)}
\ee
where $\eta(B)$  is the  weight for
nonperturbative fields, defining the  vacuum averages, e.g.
\be
<F^B_{\mu\nu}(x) \Phi^B{(x,y)} F^B_{\lambda\sigma}(y)>_B=
\frac{\hat 1}{N_c} (\delta_{\mu\lambda}\delta_{\nu\sigma}-
\delta_{\mu\sigma}\delta_{\nu\lambda})D(x-y)+\Delta_1
\ee
where $F^B_{\mu\nu},\Phi^B$ are field strength and parallel
transporter made of $B_\mu$ only; $\Delta_1$ is the full derivative
term  [4] not contributing to the string tension $\sigma$, which  is
\be
\sigma  =\frac{1}{2N_c} \int d^2 x D(x) +O(<FFFF>)
\ee

The background perturbation theory is an expansion of the last
integral in (2) in powers of $ga_\mu$ and averaging over $B_\mu$ with
the  weight  $\eta(B_\mu)$, as shown in (3). Referring the reader to
[9,10] for explicit formalism and  renormalization, we concentrate
below on the static interquark interaction at small $r$. To this end
we  consider the Wilson loop of size $r\times T$, where $T$ is large,
$T\to \infty$,  and define
\be
<W>_{B,a}= <P exp ig\int_C(B_\mu+a_\mu) dz_\mu>_{B,a}\equiv
exp\{-V(r)T\}
\ee
Expanding  (5)  in powers of $g a_\mu$, one obtains
\be
<W>=W_0+W_2+...;~~V=V_0(r)+V_2(r)+V_4(r)+,
\ee
where $V_n(r)$  corresponds to  $(g a_\mu)^n$ and can be expressed
through $D,\Delta_1$ and higher correlators [5,9];

Coming now to $V_2(r)$, describing one exchange  of perturbative
gluon in the background, one finds from the quadratic in $a_\mu$ term
in $S(B+a)$ in the background Feynman gauge the gluon Green's
function
\be
G_{\mu\nu} =-(D^2_\lambda \delta_{\mu\nu}+ 2ig F^B_{\mu\nu})^{-1},
~~D_\lambda^{ca}=\partial_\lambda \delta_{ca}+g f^{cba}
B^b_\lambda
\ee

Expanding in powers of $gF^B_{\mu\nu}, G_{\mu\nu}$ can be   written as
\be
G=-D^{-2}+D^{-2} 2ig F^B D^{-2} - D^{-2}   2ig
F^BD^{-2}2igF^BD^{-2}+...,
\ee
the first term on the r.h.s. of (8) corresponds to the spinless gluon
exchange,
propagating in the confining film covering the Wilson loop [9,10]. As
it was shown recently [11], the term $D^{-2}$ produces only week
corrections $O(r^3)$ to the usual perturbative potential at
small distances, while it corresponds to the massive
spinless  propagator with
mass $m_0$ at large distances.

 In what follows we concentrate on
the third term in (9), yielding for  $W_2^{(3)}$
 \be
 W^{(3)}_2 = T\int\frac{\alpha_s(k^2)}{\pi^2} \frac{d^3 ke^{i\vec
 k\vec r}\mu^2(k^2) }{(\vec k^2+ m^2_0)^2}= -\Delta V_2(r)T
 \ee
where we have defined,
having in mind (4)
\be
\mu^2(k^2) =6\int \frac{D(z) e^{-ikz} d^4z}{4\pi^2 z^2};
 \mu^2(0) = \frac{6\sigma N_c}{2\pi},
\ee

 From (10) one obtains the following positive contribution to the
 potential $V_2(r)$ at small $r$: (we neglect a constant term
 $O(1/m_0)$
 \be
 \Delta V_2(r)=
 \mu^2(k_{eff})\alpha_s(k_{eff}) r
 +O(r^2), ~~ r\la T_g.
 \ee
 Analysis of the integral (9) shows that  $k_{eff}\sim 1/r$, and
 therefore $\Delta V_2(r)$ is defined mostly by the short--distance
 dynamics.

   3. The analysis done   heretofore  concerns static  interquark
   potential and reveals that even at small distances NP background
   ensures some contributions which is
    encoded in the  negative  mass  squared term
   $-\mu^2$.

Applying the same NP background formalism to other processes of
interest, one would  get similar corrections of the order of
$\frac{\mu^2}{p^2}$, as was investigated in [1].

To check the selfconsistency of our results, one can find the
contribution of $\mu^2(k)$ to the correlator $D$,
\be
D(q)\sim \alpha_s(q)
\int\frac{d^4p\mu^2(p)}{(p^2+m^2_0(p))^2(q-p)^2}\sim \frac{1}{q^2},~~
q^2\to\infty
\ee
which is positive and consistent with recent lattice  data [3].
Insertion of (12) into (10) yields constant $\mu^2(p)$   at large $p$
(modulo logarithms), which implies selfconsistent NP dynamics at
small distances (large $p$).  It is worthwhile to note also that
negative sign of $\mu^2$ contribution is directly connected to the
asymptotic freedom, where the same paramagnetic term in the effective
action $S_{eff}$ [12] enters with the negative sign, and one can take
into account that $-\mu^2(x,y)\sim \frac{\delta^2S_{eff}}{\delta
a_\mu(x)\delta a_\mu(y)}. $

The author  is grateful for discussions, correspondence and very
usefull remarks to V.I.Zakharov and  useful discussions to
V.A.Novikov and V.I.Shevchenko.

The financial support of RFFI through the grants 97-02-16404 and
97-0217491  is gratefully acknowledged.

\end{document}